\documentclass[aps,prl,twocolumn,a4paper]{revtex4-1}
\usepackage{natbib}
\usepackage[T1]{fontenc}
\usepackage{amsmath}
\usepackage{graphicx}
\usepackage{amssymb}
\usepackage{siunitx}

\usepackage{ulem}
\usepackage{wasysym}

\usepackage{color}

\usepackage[dvipsnames]{xcolor}
\usepackage{subfig} 

\begin{document}

\title{Manipulation and coherence of ultra-cold atoms on a superconducting atom chip}

\author{S.~Bernon$^{*}$, H.~Hattermann$^{*}$, D.~Bothner, M.~Knufinke, P.~Weiss, F.~Jessen, D.~Cano, M.~Kemmler, R.~Kleiner, D.~Koelle and J.~Fort\'agh}

\address{CQ Center for Collective Quantum Phenomena and their Applications, 
Physikalisches Institut, Eberhard-Karls-Universit\"at T\"ubingen, 
Auf der Morgenstelle 14, D-72076 T\"ubingen, Germany}
\address{${*}$ These authors contributed equally to this work}

\begin{abstract}
\textbf{The coherence of quantum systems is crucial to quantum information processing.
 While it has been demonstrated that superconducting qubits can process quantum information at  microelectronics rates, it remains a challenge to preserve the coherence and therefore the quantum character of the information in these systems. An alternative is to share the tasks between different quantum platforms, e.g. cold atoms storing the quantum information processed by superconducting circuits. In our experiment, we characterize the coherence of superposition states of $^{87}$Rb atoms magnetically trapped on a superconducting atom-chip. We load atoms into a persistent-current trap engineered in the vicinity of an off-resonance coplanar resonator, and observe that the  coherence of hyperfine ground states is preserved for several seconds.
We show that large ensembles of a million of thermal atoms below \SI{350}{\nano\kelvin} temperature and pure Bose-Einstein condensates with $\bf{3.5\times10^5}$ atoms can be prepared  and  manipulated at the superconducting interface. This opens the path towards the rich dynamics of strong collective coupling regimes.
}
\end{abstract}
\maketitle


The quantum physics of interfaces is attracting great interest because quantum state transfer between subsystems is required for quantum measurements, quantum information processing and quantum communication \cite{Wallquist2009}. To overcome the fast decoherence of superconducting qubits, the engineering of 
 various hybrid quantum systems recently became a subject of intensive research \cite{Kubo2011,Zhu2011,Amsuss2011,Camerer2011, OConnell2010, Kalman2012}. The success of strong coupling between superconducting two-level systems and microwave cavities \cite{Wallraff2004}, the implementation of quantum algorithms with superconducting circuits \cite{DiCarlo2009, Fedorov2012, Reed2012} and the successful realization of superconducting surface traps for ultracold atoms \cite{Nirrengarten07, Mukai07, Mueller2010a} encourage the development of superconductor/cold atom hybrids. So far, some fundamental interactions between the two systems have been observed \cite{Mueller2010a,Cano08a,Kasch2010}. However, coherent coupling between the systems remains a scientific and technological challenge. Although theoretical proposals suggest using atomic ensembles as quantum memories in a hybrid quantum computer \cite{Petrosyan2008, Petrosyan2009,Verdu2009}, the trapping of atoms in the vicinity of a superconducting coplanar microwave waveguide resonator (CPR) is still required.
 
Long coherence times and state transfer are central issues for quantum information processing. In cold atomic ensembles, the fine control of inhomogeneous dephasing sources  \cite{Dudin2010,Radnaev2010} allow long storage times of a single collective excitation  \cite{Bao2012}. A similar control in chip-based trapped atomic clocks \cite{Treutlein2004}, allowed to preserve coherent states of rubidium hyperfine levels over tens of seconds  \cite{Deutsch2010}. In addition, the energy spectrum of rubidium atoms can  be used  to convert the quantum information to the near infrared, in the telecom band \cite{Radnaev2010}, where long distance quantum communication can be realized. Hybrid systems of cold atoms and superconductors are therefore very appealing for a solid state, atomic and photonic quantum interface. Nevertheless, the question how to preserve such coherence and optical  properties in the complex environment of a hybrid system needs to be solved. 
 
Here, we report on the preparation of coherent atomic samples at a superconducting interface. We load ultracold $^{87}$Rb atoms into a magnetic trap generated by a superconducting niobium thin film structure. We measure exceptionally long lifetimes of fully spin polarized states ($> \SI{200}{s}$). In either of the hyperfine ground states of rubidium, we reach the critical temperature of Bose-Einstein condensation with more than one million atoms. The coherence of the superpositions of these ground states is measured for various positions on the superconducting interface. In a self-centered persistent-current trap engineered in the vicinity of a CPR we observe a coherence time $T_2\sim \SI{7}{\second}$. This demonstrates that cold atom trap inhomogeneities can be controlled in this complex environment to a metrological level, paving the way towards long living single excitations.

\begin{figure}
\subfloat{\includegraphics[width=0.48\textwidth, natwidth=4785,natheight=2671]{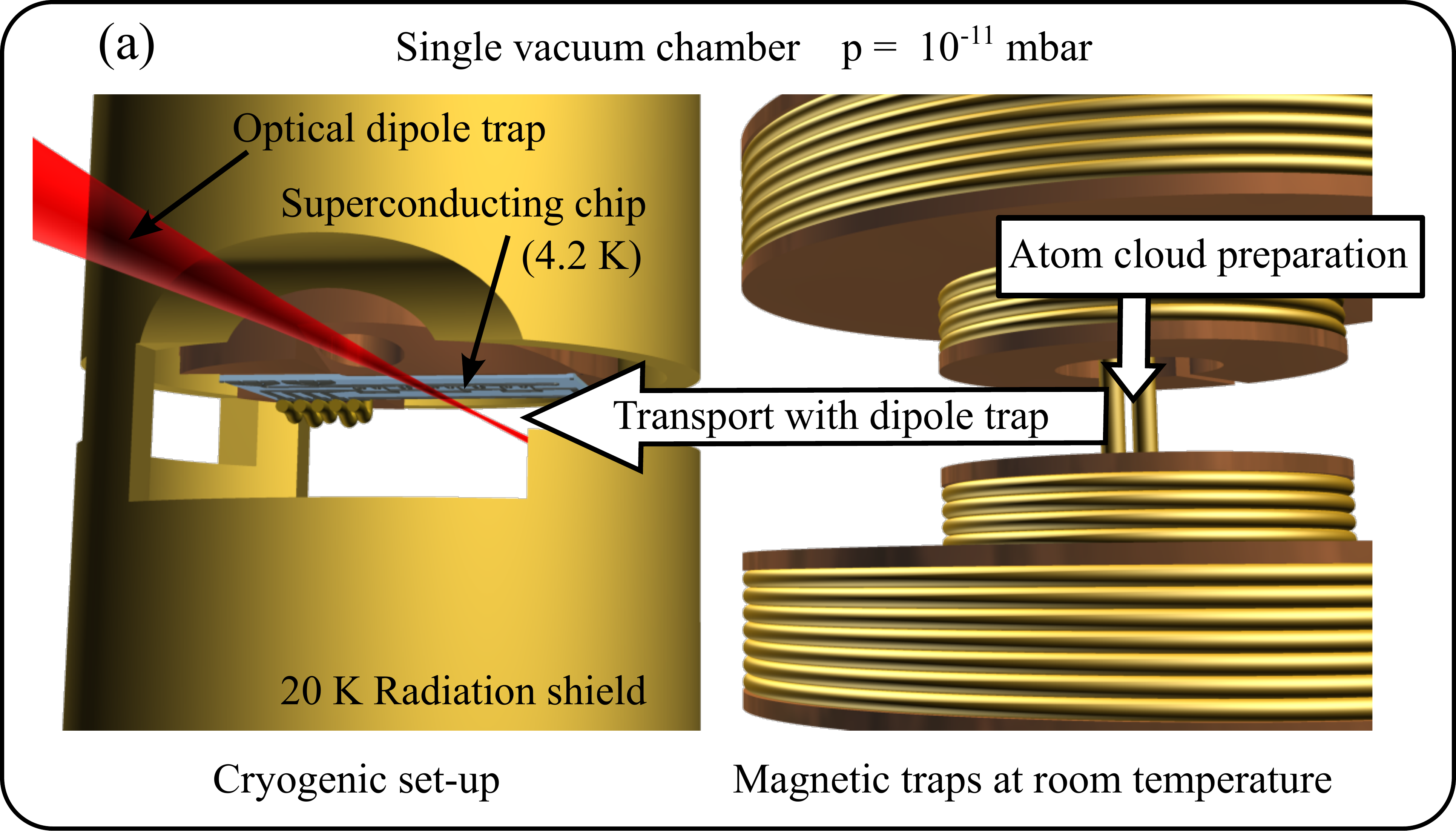}}\\
\subfloat{\includegraphics[width=0.48\textwidth]{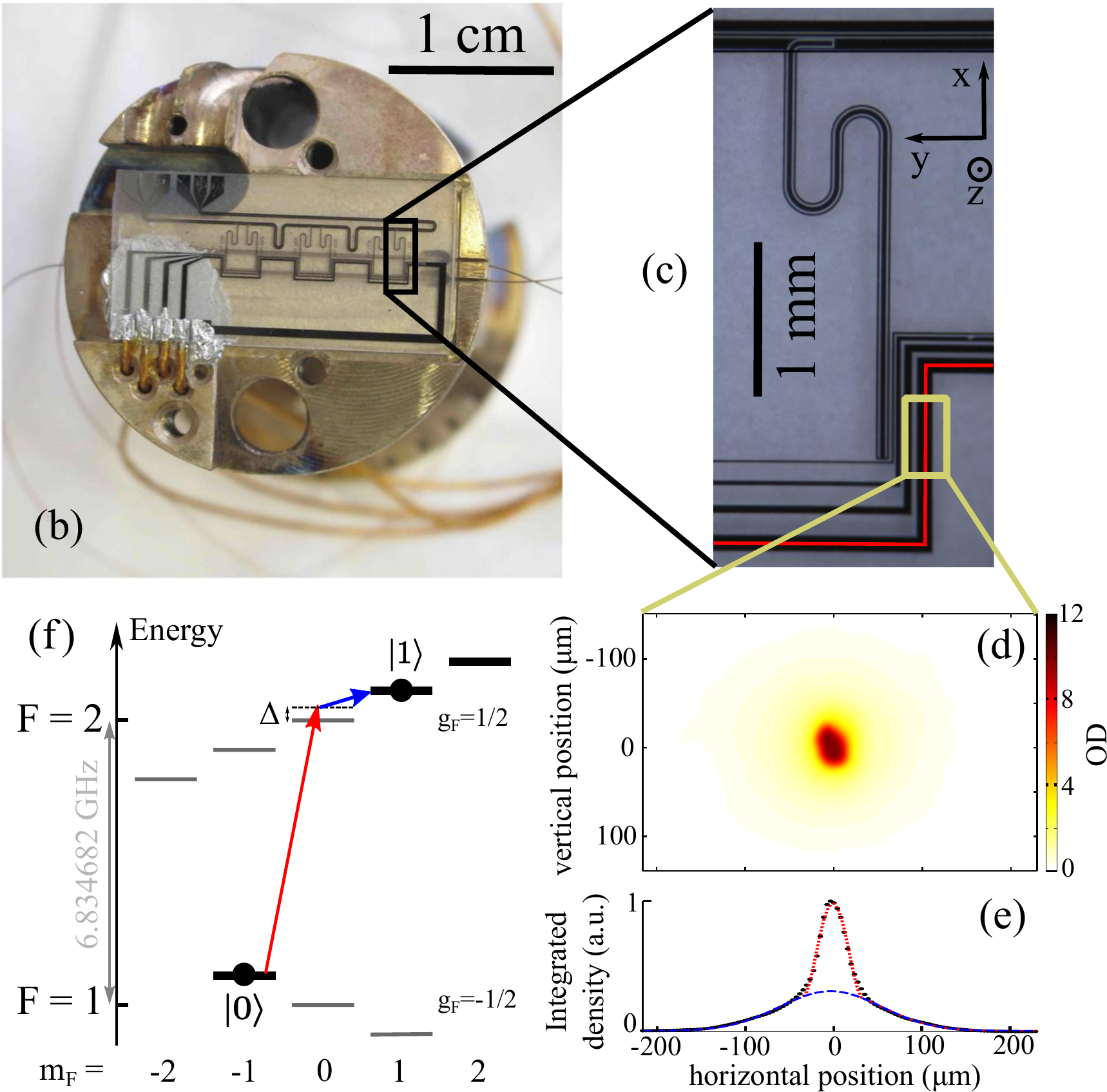}\label{fig:chipb}}

\caption{\bf{Hybrid system of ultra-cold atoms and superconductors.}}{\footnotesize{(a) In-vacuo setup (to scale): On the right side, the atoms are trapped and cooled in a room temperature environment. The left part shows the superconducting chip attached to the cryostat at $4.2$\,K and surrounded by a gold-plated radiation shield at $\sim20$\,K. The atoms are transported from one environment to the other (\SI{40}{\milli\meter} distance) by an optical tweezer. (b) Photograph of the superconducting atom-chip mounted onto an oxygen free copper holder. (c) Microscope image of the superconducting trapping structure. Visible are four $Z$-wires for trapping and a quarter-wave coplanar microwave waveguide resonator capacitively coupled to the feedline. (d) Absorption image of a Bose-Einstein condensate in state $|0\rangle$ with $N_{\rm BEC} = 3\times10^5$ atoms after 15\,ms time of flight. Color scale corresponds to optical density (OD). (e) Normalized integrated density showing the bi-modal structure of a BEC (black points). In dashed blue a fit to the thermal background and in dotted red a fit to the central Thomas Fermi profile. (f) Energy diagram of $^{87}$Rb in a magnetic field. In dark are the three magnetically trappable states. The coupling of $|F=1,m_F=-1\rangle$ ($|0\rangle$) and $|F=2,m_F=1\rangle$ ($|1\rangle$) is realized by a two photon transition.}}
\label{fig:chip}
\end{figure}

The experimental apparatus combines a cryostat (Janis ST-400, \SI{2}{W} cooling power) holding a superconducting atom-chip and a cold atom setup integrated in a single ultra high vacuum (UHV) chamber (Fig. \ref{fig:chip}(a)). The vacuum ($\sim10^{-11}$mbar) provides  excellent heat isolation between the superconducting  chip surface at a temperature of $T=\SI{4.2}{\kelvin}$ and the room temperature electromagnets that are used for the preparation of cold atomic samples \cite{Cano2011}. A copper radiation shield at $\sim$ \SI{20}{K} protects the chip from the room temperature  thermal radiation.  A slit of \SI{2}{mm} height on the shield gives optical access to an optical tweezer that transports atom clouds from the room temperature environment to the superconducting atom-chip. \\
The superconducting chip with $Z$-shaped wires (red line) and a quarter-wave CPR structure is shown in Fig.\ref{fig:chip}(b) and (c). The niobium  film structures (\SI{500}{\nano\meter} thickness) were fabricated  on monocrystalline sapphire by magnetron sputtering, optical lithography and reactive ion etching (SF$_6$). Niobium is, in our experimental conditions, a type II superconductor with a transition temperature of \SI{9.2}{\kelvin}. At \SI{4.2}{\kelvin}, the wires  carry mean current densities of up to $4\times10^6$\,A/cm$^2$, corresponding to a current of \SI{1}{\ampere} for a  wire of \SI{50}{\micro\meter} width. Supply wires (normal conducting copper) are connected by ultrasonic soldering to the niobium. The sapphire substrate is similarly soldered to the copper mount of the cryostat.  Two superconducting wires of \SI{100}{\micro\meter} diameter pass below the chip and help to maintain the longitudinal confinement of the trap (Fig.~\ref{fig:chip}(b)).

\begin{figure*}
\subfloat{\includegraphics[width=1\textwidth]{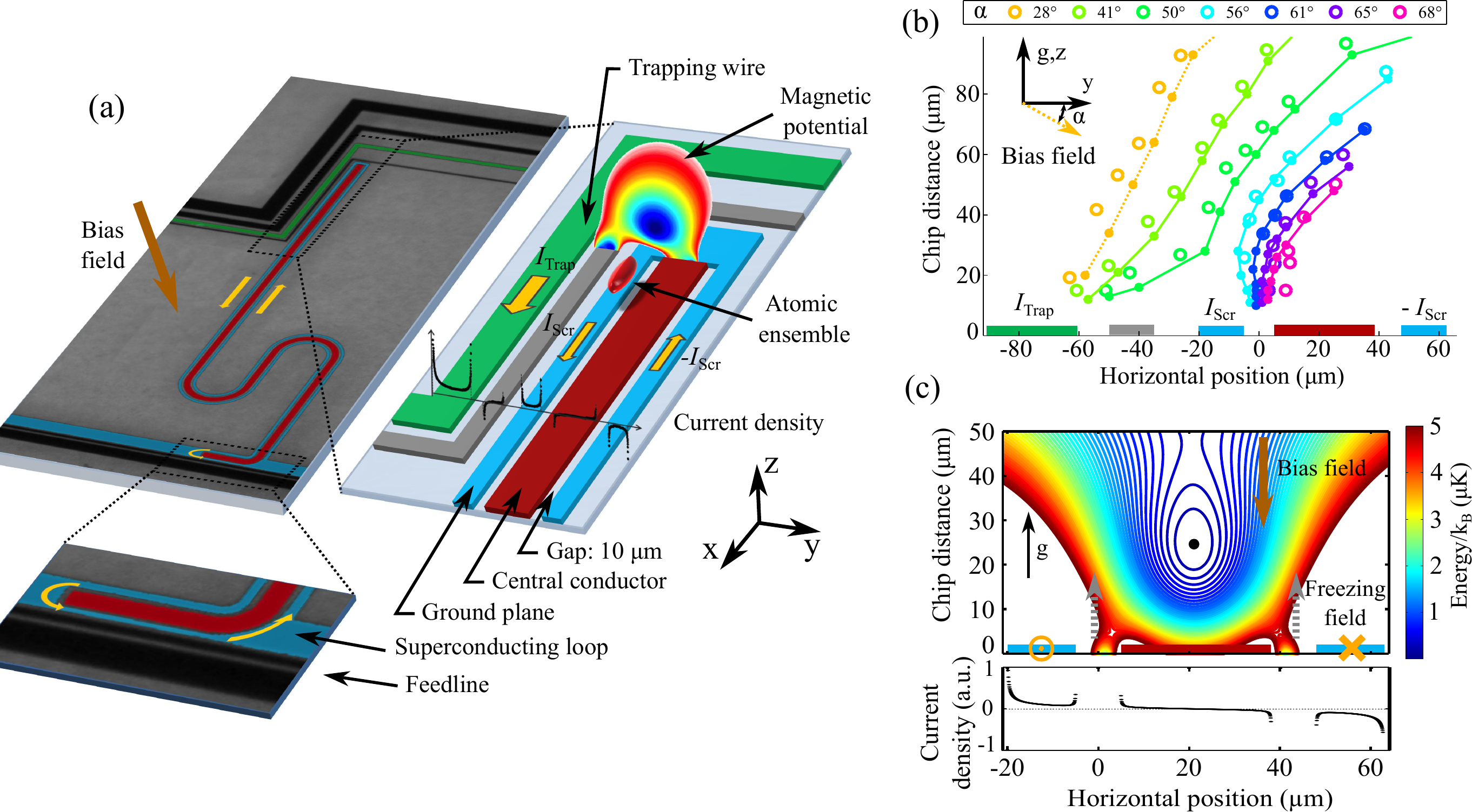} \label{fig:pos_scheme}} 

\caption{\bf{Positioning atoms close to a superconducting coplanar microwave resonator (CPR).}}{\footnotesize{(a) Scheme of the trapping of atoms in the gap of a superconducting quarter wavelength CPR. The trap inside the gap is a result of the magnetic fields generated by the current of the trapping wire $I_\text{Trap}$ and by the screening current $I_\text{Scr}$ in the ground plane. These fields cancel with an externally applied bias field. The embedded plot (black dots) is the simulated  distribution of the screening currents in the superconductor. These currents keep the flux in the superconducting loop constant and the interior of the  films field free. The transverse profile of the magnetic potential is shown in color. The dark blue corresponds to the potential minimum. (b) Position of the trap for different currents in the trapping wire and different angles $\alpha$ between the bias field and the surface of the chip. $\alpha=B_{z}^{\rm{bias}}/B_{y}^{\rm{bias}}$ is varied by changing $B_{z}^{\rm{bias}}$ with $B_{y}^{\rm{bias}}=\SI{2.3}{G}$ constant. The position of the atoms has been measured by in-situ absorption imaging (See Methods). For small angles, the trap behaves as for a normal conducting chip, i.e., when the current is reduced in the trapping wire ($I_\text{Trap}$), the trap moves towards it. For large angles, this behavior is modified and the trap is focussed into the gap between the center conductor and the ground plane of the CPR. The agreement between measurement (circles) and simulations (dotted and solid line with dots) proves that positioning of the atoms in the gap of the CPR can be facilitated by screening currents $I_\text{Scr}$ in the ground planes. The simulations \cite{Cano08,Markowsky2012} are performed with no adjustable parameter and assume a Meissner state for the superconductor. Gravity, $g$, is oriented upwards. (c) Top: Potential energy landscape of a persistent-current trap above the central conductor (red) of the CPR. This trap is generated by the superposition of a vertical homogeneous bias field $B^{\rm{bias}}\approx\SI{1.2}{G}$ and the field induced by the screening currents. To enhance the screening currents at a given bias field, a non-zero flux is trapped in the gap of the CPR during the cool down of the cryostat (Freezing field, $B^{\rm{freezing}}\approx\SI{0.8}{G}$). Isolines are separated by \SI{100}{\nano\kelvin}. Bottom: Screening current density distribution induced by the combination of bias field and freezing field.
}}
 \label{fig:pos}
\end{figure*}

The preparation of atomic clouds follows standard techniques of magneto-optical (MOT) and magnetic trapping (see Methods for details) and leads, in the room temperature environment, to a cold cloud of $5\times 10^6$ atoms at $T<\SI{1}{\micro\kelvin}$. After transfer to an optical dipole trap, the cloud is transported into the cryogenic environment to the loading position at $\sim\SI{400}{\micro\meter}$ from the superconducting chip surface. The atoms are then transferred into a  harmonic magnetic trap formed by superposing the field generated by the current $I_{\rm Trap}=\SI{0.8}{\ampere}$ driven through the largest $Z$-shaped trapping wire with an homogeneous magnetic bias field $B_\text{bias}=\SI{4}{G}$ applied along $y$ and a magnetic offset field $B_\text{off} = \SI{0.6}{G}$ applied along $x$ \cite{Fortagh07}. 
The atoms loaded are then adiabatically compressed in a trap with oscillation frequencies $\{\omega_x,\omega_y,\omega_z\}/ 2\pi =\{19, 145, 128\}$\,Hz and subsequently cooled by radio-frequency forced evaporation to temperatures of $\sim\SI{200}{\nano\kelvin}$. By changing the length of the initial MOT phase (1-10\,s), the atom number in the superconducting trap ($10^4 -  10^6$) and the final state (thermal cloud or BEC) can be conveniently controlled without affecting the temperature. 
Due to the strong cryopumping and the suppression of thermally driven magnetic field fluctuations \cite{Kasch2010}, the lifetimes of the atoms in such a surface trap are predicted to be exceptionally long \cite{Skagerstam06, Hohenester2007, Nogues2009}. For a cloud of $N_{\rm at} = 10^5$ atoms polarized in state $|0\rangle$ ($|F=1,m_F=-1\rangle$) at a density of $10^{13}$ $\text{at/cm}^{3}$ held in the compressed trap with offset field $B_\text{off}=\SI{0.6}{G}$, we measure a lifetime of more than \SI{4}{\minute}. Together with the recently measured lifetime of \SI{10}{\minute} for an experiment entirely shielded at nitrogen temperature \cite{Emmert09a}, we expect that the sub-kelvin cryoshields of dilution refrigerators  will provide excellent vacuum and thermal noise  conditions for cold atom setups.\\
We form pure Bose-Einstein condensates (BEC) with up to $3.5\times10^5$ atoms in either of the spin states $|0\rangle$ or  $|F=2, m_F=2\rangle$. The lifetime of a condensate of $4 \times 10^5$ atoms in a trap with frequencies $\{15, 72, 43\}$\,Hz is  \SI{35(5)}{\second} and is density limited by three-body collisional losses \cite{Soeding99}. Such high atom numbers are of special interest for quantum information protocols in which the collective enhancement allows one to reach an effective strong coupling regime \cite{Verdu2009}, for superradiance experiments and for the realization of an on-chip maser \cite{Henschel2010}.

In the following, we explore the properties of the cold cloud in the vicinity of the superconducting CPR. We first consider the positioning of  atoms in the close vicinity of the CPR where we additionally engineer a persistent-current trap. As a further step, we study the coherence of atomic superposition states in different positions of the CPR mode volume. In this report, the resonator is, by design, off  resonance from the atomic transition and should therefore not affect the internal state dynamics of the sample.

\paragraph{\textbf{Positioning of atomic clouds into a CPR}}

\begin{figure*}
\subfloat{\includegraphics[width=0.65\textwidth]{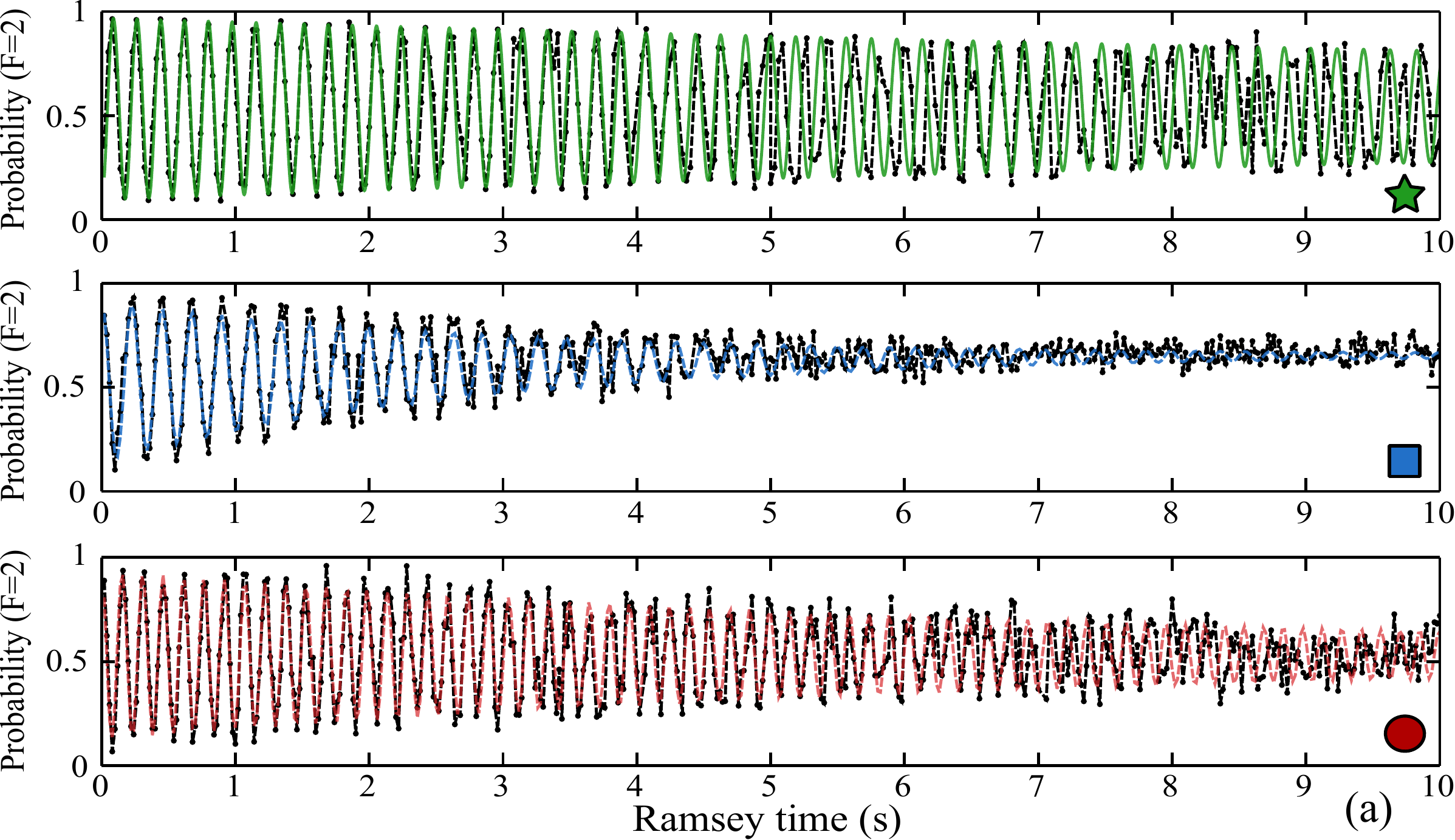}\label{fig:ramsey_plot}} \hspace{0.3cm}
\subfloat{\includegraphics[width=0.32\textwidth]{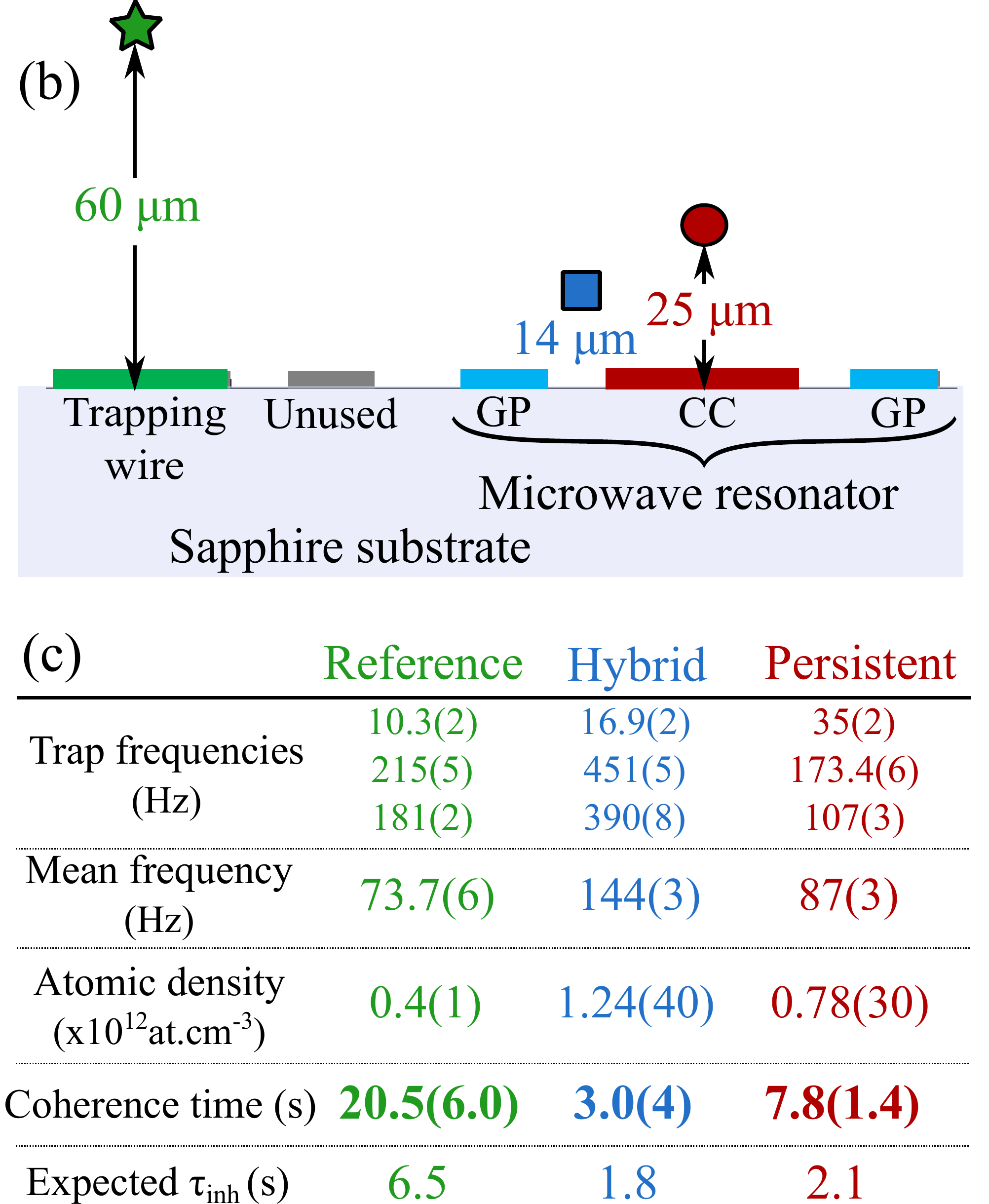}} 
\caption{\bf{Atomic coherence in a  superconducting coplanar microwave resonator (CPR).}}{\footnotesize{(a) Ramsey fringes measured in the time domain for different trapping positions on the superconducting atom-chip. From top to bottom, the trapping positions correspond to the reference trap (green star in (b)), the hybrid trap (blue square in (b)) and the persistent-current  trap (red disk in (b)). The coherence time obtained for the three sets are respectively \SI{20.5}{\second}, \SI{3}{\second} and \SI{7.8}{\second} that indicate collision induced spin self rephasing effects \cite{Deutsch2010} (see main text and Methods) (b) Positions of the different traps below the superconducting chip. The reference trap (green star) is situated \SI{60}{\micro\meter} straight below the trapping wire. The hybrid trap, that is generated by applied and induced currents, stands \SI{14}{\micro\meter} below the \SI{10}{\micro\meter} wide gap of the CPR. The persistent-current trap is  \SI{25}{\micro\meter} below the central conductor (CC) of the CPR that has a width of \SI{33}{\micro\meter}. GP stands for ground planes. Horizontal and vertical axes are to scale. (c) Table of the experimental parameters for the three measurements shown in (a).  The density quoted is the mean atomic density. The mean frequency is the geometrical average.  $\tau_{\rm inh}=\sqrt{2}/\Delta_0$ is the coherence time expected for the corresponding residual frequency inhomogeneity $\Delta_0$ (see Methods).} }

 \label{fig:Ramsey}
\end{figure*}

In a CPR, the electromagnetic fields are concentrated in the gaps between the ground planes and the central conductor. To maximize the atom-cavity coupling, atoms need to be positioned in the close vicinity of these gaps.  
Nevertheless, for trap-wire distances smaller than the wire width, the trapping parameters are significantly affected  by the Meissner-Ochsenfeld effect (MOE) \cite{Cano08a}. To limit such deformations in the vicinity of the CPR, the design shown in Fig.\ref{fig:chip}(c) includes four $Z$-shaped wires with widths ranging from  \SI{100}{\micro\meter}  to \SI{15}{\micro\meter}. Starting from the largest $Z$-wire trap, the atoms are first horizontally transferred to the third $Z$-wire trap. The fourth and last wire (\SI{15}{\micro\meter} width) 
revealed to be unnecessary. 
At \SI{100}{\micro\meter} below the third trapping wire, the trap needs to be rotated towards the CPR gap.  Two parameters are classically used to manipulate atoms on a chip; the value of the  trapping wire current modifies the trap-wire distance and the direction of the bias field rotates the trap around the trapping wire. Because of the MOE, positioning that is  straight forward for a normal conducting chip, is more subtle for a superconducting atom-chip. In the present case, we take advantage of the conservation of magnetic flux in superconducting loops to directly guide the atoms below the gap of the CPR.

The design of the quarter-wave CPR includes a superconducting loop formed by the ground planes of the resonator (blue in Fig. \ref{fig:pos}(a)). When a field perpendicular to the substrate is applied, such as the bias field, a screening current is induced in the ground planes and ensures the conservation of the magnetic flux in this superconducting loop. Such a current, that circulates just next to the gap, generates a magnetic field profile that guides the atoms  into the gap. At distances comparable to the width of the ground planes, such guiding is further enhanced by the MOE that focusses magnetic field lines and generates magnetic gradients that center the cloud in the gap. The guiding of atoms into the gap is observed by in-situ measurements of the position of the atomic cloud for different bias field orientations (angle $\alpha$ in Fig. \ref{fig:pos}(b)) 
and different currents in the wire (Fig. \ref{fig:pos}(b)). For each experimental point, the cloud is first brought to a trap-wire distance of $\sim\SI{100}{\micro\meter}$, rotated to the angle $\alpha$ and then moved to the desired trapping current. The measured position $(y,z)$ agrees well with a 2D simulation  of the London equations \cite{Cano08}, that includes gravity and the conservation of flux in the resonator loop (see Methods). The simulations are performed without free parameters.
As experimentally observed and consistent with our model, we note that the two gaps of the CPR are not equivalent. Due to the opposite direction of the current in the ground planes and the orientation of the bias field, only the  closest gap  to the trapping wire can be accessed. We call this trap, resulting from applied and induced currents, a hybrid trap.  

A further advantage in using superconductors to manipulate cold atoms, is the possibility to engineer traps induced by persistent currents. These persistent-current traps \cite{Mukai07}  render the direct injection of currents on the chip unnecessary and therefore suppress the related source of noise. In our geometry, such a trap can be generated directly below the CPR. Here we demonstrate a self-centered trap generated by a single homogeneous bias field and its induced  screening currents. By trapping flux in the resonator loop with a freezing field applied during the cooling of the cryostat, the magnitude of the screening currents can be  controlled independently from the applied bias field. This experimentally realized trap is simulated in Fig. \ref{fig:pos}(c) where both the resulting potential and the current densities are depicted. The simulation presented includes the distortion due to the MOE, the conservation of magnetic flux in the loop and the effect of vortices trapped inside the film during the cooling  of the cryostat \cite{Stan2004}. The agreement of position and trap frequencies between the experiment and our simple model requires an adjustment of the input parameters by less than $20\%$. For a small BEC in the trap of Fig. \ref{fig:pos}(c) with oscillation frequencies $\{35,173,107\}$ Hz, we measure a lifetime of \SI{10(4)}{\second}, probably limited by the residual technical noise and the shallow trap depth of $\sim \SI{320}{\nano\kelvin}$.

\paragraph{\textbf{Coherence in a superconducting CPR}}
To study the coherence lifetime of atomic ensembles in the close vicinity of a CPR, we perform Ramsey measurements at different positions below the superconducting chip. These measurements compare the coherence of a trapped atomic ensemble in a superposition of the states $|0\rangle$ and $|1\rangle$ to a high stability \SI{10}{\mega\hertz} reference oscillator that has a short and long term frequency stability $\Delta f/f< 5\times10^{-12}$. These two states are chosen for the low sensitivity of their transition frequency to magnetic inhomogeneities at a magnetic offset field of 3.228 G (see Methods) \cite{Harber2002}. 
The atomic cloud is first prepared in a thermal and pure state of $|0\rangle$. The initialization of the coherent superposition is realized by a $\pi/2$~two-photon excitation that starts the interferometric sequence. After a variable waiting time $T_R$, the interferometer is closed by a second $\pi/2$~pulse. The populations $N_0(T_R)$ and $N_1(T_R)$ in respectively $|F=1\rangle$ and $|F=2\rangle$ are consecutively read out by state selective absorption imaging (See Methods), and the resulting probability of $|F=2\rangle$:~$N_1(T_R)/(N_0(T_R)+N_1(T_R))$ is displayed in Fig. \ref{fig:Ramsey}(a).\\
To study the capabilities of our setup, we first describe a reference measurement that is performed far from the CPR: \SI{60}{\micro\meter} below the third trapping wire. This measurement is conducted with  $1.9(4)\times 10^4$ atoms at a temperature of \SI{245(30)}{\nano\kelvin} in a magnetic trap with oscillation frequencies $\{10,215,181\}$\,Hz, corresponding to a mean density (in units of $10^{12}$\,at/cm$^{3}$) $\overline{n}\approx 0.4(10)$ (green star in Fig. \ref{fig:Ramsey}(a)). In this configuration, the coherence time is $T_2=20.5(6.0)$ s ($1/e$ exponential decay time). Due to magnetic noise in this non-shielded apparatus and drifts over the length of the scan (several hours), the phase starts to be lost for $T_R>\SI{5}{\second}$. This decay time $T_2$ exceeds the time of $\SI{6}{\second}$ predicted by the residual trap inhomogeneities $\Delta_0/2\pi\approx \SI{0.04}{\hertz}$ (see Methods, \cite{Rosenbusch2009}). This indicates that we entered the spin self-rephasing regime \cite{Deutsch2010}, where the identical spin rotation rate $\omega_{\rm ex}/2\pi=\SI{3}{\hertz}$ dominates  both $\Delta_0$ and the rate of lateral elastic collisions $\gamma_{\rm c}=\SI{1}{\per\second}$. This regime can be understood as a continuous spin-echo process triggered by forward atomic collisions.\\ 
The second trap under study is the hybrid trap, situated \SI{14}{\micro\meter} below the gap of the CPR. As mentioned before, this position is particularly privileged for the perspective of strong atom-cavity coupling. The Ramsey fringes shown by blue squares in Fig. \ref{fig:Ramsey}(a) were obtained for $1.25(40)\times 10^4$ atoms at a temperature of \SI{320(20)}{\nano\kelvin} held in a confined trap $\{16.9,451,390\}$\,Hz, corresponding to $\overline{n} \approx 1.24(40)$. 
In the present situation, the coherence time is $T_2=\SI{3.0(4)}{\second}$ and is mainly limited by the atom loss during $T_R$ that has a decay time of \SI{3.2(2)}{\second}. This loss is probably due to spin exchanging collisions from atoms in state $|1\rangle$ towards the states  $|F=2, m_F=0\rangle$ (untrapped) and $|F=2, m_F=2\rangle$ (trapped) \cite{Egorov2011}. 
One typical signature of this loss channel is the asymmetry of the probability for large $T_R$. These large collisional losses are due to the strong confinement of this trap. This is the drawback of the guiding mechanism that involves strong magnetic gradients generated by the close-by screening currents. The fringes presented in figure \ref{fig:Ramsey}(a) result from an optimization of the temperature that is high enough to minimize collisional losses and low enough to enter the spin-self rephasing regime  ($\omega_{\rm ex}/2\pi=\SI{9.5}{\hertz}$, $\gamma_{\rm c}=\SI{3.6}{\per\second}$,  $\Delta_0/2\pi=\SI{0.12}{\hertz}$). \\
The last position studied corresponds to the persistent-current trap previously mentioned. In this trap the measurement was performed with $1.25(40)\times 10^4$ atoms at \SI{158(23)}{\nano\kelvin}, corresponding to $\overline{n} \approx 0.78(30)$ and to the related spin self rephasing parameters $\omega_{\rm ex}/2\pi=\SI{6}{\hertz}$, $\gamma_{\rm c}=\SI{1.6}{\per\second}$,  $\Delta_0/2\pi=\SI{0.1}{\hertz}$. The coherence lifetime obtained is $T_2=\SI{7.8(14)}{\second}$ and is mainly limited, in this shallow trap ($\sim \SI{320}{\nano\kelvin}$), by the atom loss decay time \SI{4.8(3)}{\second}.\\
The three measurements are compared in Fig. \ref{fig:Ramsey}(c). It shows that the trap deformation induced by the superconducting CPR results in an increase of the trap frequencies which strongly impact the coherence of the atomic cloud. It also proves that this influence can be engineered to reach performances comparable to the state of the art of metrological experiments \cite{Deutsch2010}.

In conclusion, we have demonstrated that hybrid systems of cold atoms and superconductors are now sufficiently mature to produce excellent experimental conditions for the study of the quantum properties of such an interface. The production of large thermal clouds and BECs of a million atoms opens the path to the strong collective coupling to a CPR, and to the transfer of quantum information between the two systems. We also demonstrated that, in the vicinity of a superconducting CPR, magnetic traps could be efficiently engineered to produce robust and controllable conditions. The measurement of coherence lifetimes of more than \SI{3}{\second} in this new type of environment gives strong hope that cold atoms could be used as a quantum memory for superconducting devices. We also point out that the motion of particles plays  an interesting role here. While this is usually considered to be a source of decoherence and prevents, for example, the use of spin-echo techniques, it might actually be turned to preserve the coherence of a quantum memory.

\paragraph{\textbf{\large{Acknowledgements:}}}
We would like to thank Thomas Udem from the MPQ Munich as well as Max Kahmann and Ekkehard Peik from the PTB Braunschweig for the loan of the reference oscillators.
This work was supported by the Deutsche Forschungsgemeinschaft (SFB TR 21) and ERC (Socathes). H.H. and D.B.  acknowledge support from the Evangelisches Studienwerk e.V.Villigst. M.Kn. and M.Ke.  acknowledge support from the Carl Zeiss Stiftung.

\paragraph{\textbf{\large{Author contributions}}}
D.K., R.K., J.F., S.B., H.H., M.Kn. and M.Ke. designed and mounted the experiment. D.B. ,M.Kn., M.Ke. and H.H. designed and fabricated the superconducting chip.  S.B., H.H. and P.W. carried out the experiments and analyzed the data. S.B., H.H., D.C., M.K. and D.B. made the numerical simulations. D.K., R.K. and J.F. supervised the project. S.B., H.H. and J.F. edited the manuscript. All authors discussed the results and commented on the manuscript. 


\paragraph{\textbf{\large{Methods:}}}
\paragraph{\textbf{{Atom cloud preparation:}}}
\footnotesize{
The magneto-optical trap (MOT) is loaded from a 2D-MOT. For a loading time of \SI{6}{s} , the MOT contains $\sim10^9$ $^{87}$Rb atoms at a temperature of {$\sim$}\SI{200}{\micro \kelvin}. With this method, we do not observe perturbation of the background pressure. After an optical molasses, the atomic cloud is optically pumped into one of the two hyperfine ground states $|0\rangle$ or $|F=2, m_F=2\rangle$ (total angular momentum states) and is transferred through a magnetic quadrupole into a harmonic Ioffe-Pritchard type trap. It is further cooled by forced RF evaporation to a temperature of $\sim$\SI{1.5}{\micro \kelvin}, slightly above the BEC transition. The remaining $5\times 10^6$ atoms are loaded into an optical tweezer ($\lambda = 1064$\,nm laser, $P=500$\,mW, focused to $w_0 = $ \SI{25}{\micro \meter} beam waist) and  transported without significant loss or heating over a distance of \SI{40}{mm} to the superconducting chip. During the optical transfer, a quantization field of 350 mG along $x$ is applied to maintain the polarization of the sample.}

\paragraph{\textbf{{Magnetic field calibration:}}}
\footnotesize{At the position of the atomic cloud, below the superconducting chip, the magnetic field is controlled by 3 orthogonal pairs of coils that allow to independently control the 3 directions of space. The calibration of the residual field and the field generated by the coils is realized in-situ by microwave spectroscopy. To that purpose, the atoms are prepared in $|0\rangle$ and the magnetic field is set to the desired field of study. After all Eddy currents have  damped out, the atoms are released from the optical dipole trap. The microwave transition considered is $|0\rangle$ to $|F=2, m_F = 0\rangle$. In the absence of a magnetic field it has a frequency of \SI{6.834682610}{\giga\hertz} and a first order Zeeman sensitivity of \SI{+700}{\kilo\hertz}/G. In a first coarse step, the absolute field is reduced to below 5 mG. In a second fine step, each pair of coils is switched on in a row to generate a frequency shift of approximately \SI{200}{\kilo\hertz}. This shift is then measured with an accuracy of $\pm$\SI{500}{\hertz}. This way, each pair of coils is in-situ and independently calibrated (residual contribution from other pairs below 1\permil) and the residual field is measured to below  1 mG while the coils are calibrated with an accuracy better than 3\permil.\\
 To avoid perturbation by the MOE of the nearby superconductor, this calibration is done with the cryostat at $T\approx\SI{10}{\kelvin}$.}

\paragraph{\textbf{{Imaging and measurement of position:}}}
\footnotesize{The atoms are observed by absorption imaging with a variable time of flight. Due to Eddy currents in the mechanical system, the measured atom number is not absolute and depends on the time of flight (TOF). For TOF$<\SI{10}{\milli\second}$, the calibration of the detection of the state $|0\rangle$ with respect to the state  $|1\rangle$ is obtained by minimizing the variance of the total atom number detected  over the length of the scan. For TOF>\SI{10}{\milli\second}, atom numbers stay constant, showing that Eddy currents are no more an issue at the corresponding distances. The absolute calibration of the atom number is obtained from the critical temperature of  Bose-Einstein condensation. This calibration is the main source of uncertainty of the atom numbers.  Therefore, the atom number uncertainty quoted in the text do not represent shot to shot fluctuations.
 \textit{In-situ}, the position of the atomic cloud is measured along the three directions of space by two reflection imaging systems that are aligned along $x$ and $y$. The calibration of distances is realized by time of flight of the magnetically insensitive state $|F=2, m_F = 0\rangle$. To avoid spurious effect, the chip is uncoated.
 
\paragraph{\textbf{{Magnetic field calculations:}}}
\footnotesize{To calculate the current densities in the superconductor and the subsequent atomic trap deformation, we solved the London equations using the 2D algorithm described in \cite{Cano08}. This treatment is particularly valid in the trapping region where thin films are parallel to each other. The conservation of flux in the superconducting loop of the resonator is further taken into account by imposing net currents  in the grounds of the resonator. The influence of an homogeneous density of vortices pointing along $u_z$ is  modeled  by the superposition of an homogeneous field along $u_z$ (undeformed) and the opposite field deformed by the superconducting structure in a pure Meissner state. In all simulations, the quantization of flux  is neglected ($\Phi_0/\Phi\approx 5 \times 10^{-5}$). The simulations presented in Fig. \ref{fig:pos} have been confirmed using a 3D simulator (3D-MLSI \cite{Khapaev03}). The effect of gravity is included in all simulations.}

\paragraph{\textbf{Details on the measurement of coherence lifetime:}}
\footnotesize{The measurement of the $T_2$ time of the atomic qubit formed by the state $|0\rangle$ and $|1\rangle$ is realized by a Ramsey type experiment. 
 As shown in Fig. \ref{fig:chip}(f), the two states are coupled via a two photon transition involving a microwave photon at $f_{\rm MW} = \SI{6.83337816}{\giga\hertz}$ and an RF photon at $f_{\rm RF} = \SI{1.3}{\mega\hertz}$. Both frequencies are generated by commercial synthesizer phase locked to a high stability \SI{10}{\mega\hertz} quartz oscillator (Oscilloquartz, 8607-BHM15), and their sum is frequency detuned form the atomic transition by $\Delta_{\rm{R}}/2\pi$. For all the measurement presented, the microwave with power $P_{\rm{MW}}$ is radiated by an helicoidal antenna situated outside the vacuum chamber at a distance from the atoms of 20 cm. The radio frequency with power $P_{\rm{RF}}$  is  coupled on the chip to the largest $Z$-wire. The Rabi frequency $\Omega_{R}$  obtained in each situation is summarized in table \ref{tab:parameters}. Figure \ref{fig:Breit} shows the differential frequency shift of the qubit transition that is well approximated by $\Delta\nu(\mathbf{r})=\nu_{|1\rangle}-\nu_{|0\rangle}=\Delta\nu_0+\beta (B(\mathbf{r})-B_0)^2$, with $\Delta\nu_0=\SI{4.4973}{\kilo\hertz}$, $B_0=\SI{3.228917(3)}{G}$\cite{Deutsch2010} and $\beta=\SI{431.36}{\hertz/G^2}$ \cite{Harber2002}. At the magic offset field $B_0$, this shift is first order insensitive to the  magnetic field. The sensitivity of the coherence time to the magnetic inhomogeneities of the trap is therefore highly reduced. The measurements presented in Fig. \ref{fig:ramsey_plot}} are performed with an offset field $B_{\rm{off}}$ slightly lower than $B_0$. This configuration is known as the mutual compensation scheme \cite{Harber2002,Rosenbusch2009}, which allows to compensate the negative collisional shift $\Delta_{\rm c}(\mathbf{r})/2\pi= -0.4\,n(\mathbf{r})/10^{12}$ Hz by the positive magnetic shift $\Delta_{\rm B}(\mathbf{r})=2\pi\Delta\nu(\mathbf{r})$. In such conditions, an optimum residual radial frequency homogeneity $\Delta_0=\sqrt{\langle \left( \Delta_{\rm c}(\mathbf{r}) +\Delta_{\rm B}(\mathbf{r})\right)^2 \rangle- \langle  \Delta_{\rm c}(\mathbf{r}) +\Delta_{\rm B}(\mathbf{r}) \rangle^2}$  is obtained for an optimal offset field $B_1^{\rm opt}$ that depends on the number of particles, the temperature and the geometry of the trap.  In table \ref{tab:parameters}, we give the optimum value of $\Delta_0$ for each experimental configuration. In the abscence of spin-rephasing, such inhomogeneities should result in a decay of the Ramsey contrast with a time constant $\tau_{\rm inh}=\sqrt{2}/\Delta_0$\cite{Rosenbusch2009}.

\begin{figure}
\includegraphics[width=0.30\textwidth]{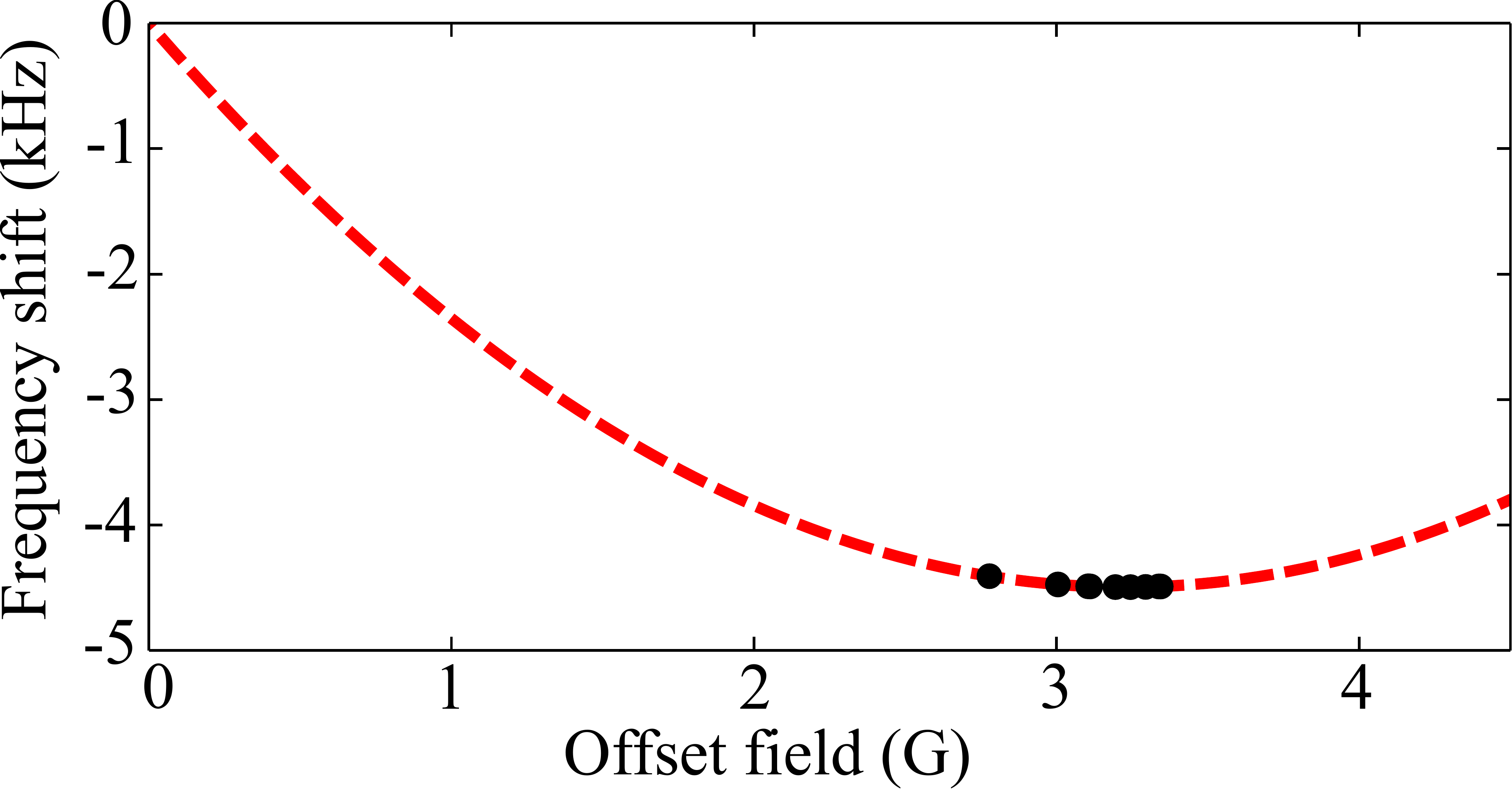}
\caption{\bf{Magnetic field dependence of the differential frequency shift of the transition $|0\rangle$ to $|1\rangle$.}}{\footnotesize{The black points are experimental data. The red dashed line is the prediction given by the Breit-Rabi formula \cite{Breit1931}.} }
 \label{fig:Breit}
\end{figure}

\begin{table}[htdp]
\caption{Experimental parameters of the measurement shown in Fig. \ref{fig:ramsey_plot}.}
\begin{center}
\begin{tabular}{|c|c|c|c|}
\hline
Properties & Reference & Hybrid & Persistent\\
\hline
$P_{\rm MW}$ (dBm) & 19 & -10 & -14\\ 
$P_{\rm RF}$ (dBm) &  -2 & 11.5 & 11.5\\ 
$\Delta_R/2\pi$ (Hz) & 5.7 & 4.4 & 6.5\\
$\Omega_R/2\pi$ (Hz) & 432 & 416  & 179\\
$B_1$ (G) & 3.193(4) & 3.086(8) & 3.20(3)  \\
$B_1^{\rm opt}$ (G) & 3.197 & 3.17 & 3.17  \\
$\Delta_0/2\pi$ (Hz) & 0.04 & 0.12 & 0.1\\
$\tau_{\rm inh}$ (s) & 6.5 & 1.8 & 2.1\\

\hline
\end{tabular}
\end{center}
\label{tab:parameters}
\end{table}%

\end{document}